\documentclass[english,a4paper]{article}
\usepackage[T1]{fontenc}
\usepackage[cp1250]{inputenc}
\usepackage{amssymb}
\usepackage{hyperref}
\usepackage{amsfonts}
\usepackage{graphicx}
\usepackage{revsymb}
\usepackage{dcolumn}
\usepackage{bm}

\newtheorem{algorithm}{Algorithm}
\newtheorem{remark}{Remark}

\def\openone{\leavevmode\hbox{\small1\kern-3.3pt\normalsize1}}
\begin{document}

\title{Newton algorithm for Hamiltonian characterization in quantum control}

\author{M. Ndong, J. Salomon\footnote{CEREMADE, UMR CNRS 7534,
  Universit\'e Paris-Dauphine, Place du Mar\'echal de Lattre de Tassigny,
  F-75775 Paris cedex 16, France} and D. Sugny\footnote{Laboratoire Interdisciplinaire
Carnot de Bourgogne (ICB), UMR 5209 CNRS-Universit\'e de
Bourgogne, 9 Av. A. Savary, BP 47 870, F-21078 DIJON Cedex,
FRANCE, dominique.sugny@u-bourgogne.fr}}
\maketitle

\begin{abstract}
We propose a Newton algorithm to characterize the Hamiltonian of a quantum system interacting
with a given laser field. The algorithm is based on the assumption that the evolution operator of the system is perfectly known at a fixed time. The computational scheme uses the Crank-Nicholson approximation to explicitly determine the derivatives of the propagator with respect to the Hamiltonians of the system. In order to globalize this algorithm, we use a continuation
method that improves its convergence properties. This technique is applied to a two-level
quantum system and to a molecular one with a double-well potential.
The numerical tests show that accurate estimates of the unknown parameters are obtained
in some cases. We discuss the numerical limits of the algorithm in terms of basin of
convergence and non uniqueness of the solution.
\end{abstract}



\section{Introduction}
\label{sec:intro}

The control of quantum systems by means of intense laser pulses has been a topic of increasing interest in
the past decades \cite{rice,shapiro,reviewQC}. It has now become a well-recognized field of research with
applications ranging from chemistry and physics to material science and nanotechnology.
In this context, several advances have been achieved extending, on the theoretical side, from the discovery
of elementary basic mechanisms of field-induced dynamics \cite{rice,shapiro} to optimal control algorithms
\cite{lapertglaser,m2an,contlett,num,pont,kosloff,maday,zhu}. The recent progress of numerical optimization
techniques has made possible the design of control fields able to manipulate quantum systems of growing complexity.
However, to be efficient, such open-loop control methods require the accurate knowledge of the dynamics of
the system \cite{gershenzon,zhang,assemat,turinici,viellard,pentlehner}.
In this framework, on the basis of different measurements of the system from different quantum states, quantum process tomography (QPT) is a set of techniques allowing to identify the dynamical map, which relates the initial states to the final ones (see, e.g. \cite{QPT1,QPT2,QPT3} and references therein). Such data can then be used to characterize the Hamiltonian or other parameters of the system \cite{schirmer,geremia,feng,geremia2,lebris,alis,shenvi,tadi,Laurent}. In this paper, we assume that a full and ideal QPT is available on the system under concern. This leads to a perfect knowledge of the evolution operator at a given fixed time. Starting from this identification, the goal of our approach is thus to determine the Hamiltonians of the system, i.e. the field-free Hamiltonian and the interaction operator. For that purpose, we also assume that a known time-dependent external field is applied to the system during the identification process. The framework of the method can be described more precisely as follows. We study a quantum system whose dynamics is ruled by the time-dependent Schr\"odinger equation. The control field is known at any time,
only one control field being used. From the final evolution operator of the system, we propose a Newton algorithm to compute the Hamiltonian of the system and the operator describing the interaction with the field. Note that the computational scheme uses the Crank-Nicholson approximation to explicitly determine the derivatives of the propagator with respect to the Hamiltonians of the system. A continuation method on the target state is used to improve the convergence properties of the algorithm.
We also discuss the singularities that can occur in our method.

The paper is organized as follows. In Sec. \ref{sec:method}, we present the theoretical framework and we introduce
the numerical procedures that will be used. Section \ref{sec:appl} is devoted to the application of this approach
to two basic quantum systems, a two-level system and a molecular one with a double-well potential. Conclusion and
prospective views are given in Sec. \ref{conc}. Technical computations about the singularities of the Newton algorithm are reported in the  \ref{app}.

\section{Theoretical framework}
\label{sec:method}
We start with a presentation of the model system and we introduce the different numerical algorithms.
\subsection{Hamiltonian characterization problem }
\label{subsec:setproblem}
Let us consider a quantum system interacting with an electromagnetic
field, whose dynamics is governed by the time-dependent Schr\"odinger
equation. We consider one (scalar) control field $E(t)$ but the method could be generalized to several fields.
We assume that the
field enters linearly in the Hamiltonian through a dipole coupling, the non linear interaction being neglected \cite{sakai,tehini}.
In this framework, the evolution equation reads:
\begin{equation}
  \left\{
  \begin{array}{lll}
    i\frac{\partial}{\partial t} U(t) &= &\big[H_0 +E(t)H_1\big]U(t) \\
                               U(t=0) &=& U_0
  \end{array}
    \right. \,,
  \label{eq:dynU}
\end{equation}
where $U(t)$ is an unitary operator that describes the state of the system at time $t$
and $H_0$ and $H_1$ are the field-free and the coupling Hamiltonians
respectively. Note that the formalism presented here can be extended to pure or mixed quantum states.
Without loss of generality, we assume that the entries on the diagonal of $H_1$ are zero,
which is the case for a dipole coupling. The matrix $U_0$ is the initial condition of the dynamical system. Units such that $\hbar=1$ are used throughout the paper. The Hamiltonian characterization is an inverse engineering control problem where for
given initial and targets states, $U_0$ and $U_\mathrm{tar}$ respectively and a known electromagnetic field,
$E(t)$, the goal
consists in identifying the pair ($H_0,H_1$) such that:
\begin{equation}
  U(t_f) = U_\mathrm{tar} \,,
  \label{eq:UtfUtar}
\end{equation}
where $t_f$ is the fixed final time of the control. We introduce the mapping
\begin{eqnarray}
  \varphi:  && \mathcal{S}^{N_d,N_d} \times \mathcal{S}_0^{N_d,N_d} \rightarrow  \mathcal{U}^{N_d,N_d} \nonumber \\
            && \qquad (H_0,H_1)  \quad \mapsto  \quad U(t_f),
  \label{eq:mapphi}
\end{eqnarray}
where, $\mathcal{S}^{N_d,N_d}$ and $\mathcal{S}_0^{N_d,N_d}$ are the sets of hermitian
Hamiltonians and hermitian Hamiltonians with diagonal entries equal to zero, respectively. The set $\mathcal{U}^{N_d,N_d}$
denotes the set of unitary matrices of size $N_d\times N_d$.
From a mathematical point of view, the  operator characterization control problem is
 equivalent to the investigation of the surjectivity of $\varphi$. This means determining the pre-image of $U(t_f)$ by the mapping $\varphi$. Here, we focus on the case of real Hamiltonians, which covers a wide range of applications.

The Newton method to solve the corresponding equations requires to differentiate the function under consideration. For the sake of completeness, we recall that in the case of
 the mapping $\varphi$ defined in
 Eq.~(\ref{eq:mapphi}), the differential is given by:
\begin{eqnarray}
  d\varphi(H_0,H_1):  && \mathcal{S}^{N_d,N_d} \times \mathcal{S}^{N_d,N_d} \rightarrow  \mathcal{A}_{H_0,H_1} \nonumber \\
            && \qquad  (\delta H_0,\delta H_1)  \,\, \,\mapsto  \, \delta U(t_f),
  \label{eq:mapdphi}
\end{eqnarray}
where, for a given pair ($H_0,H_1)$, $\mathcal{A}_{H_0,H_1}$ is the tangent space of $\mathcal{U}^{N_d,N_d}$ at $U=U(t_f)$, the final state associated with the trajectory corresponding to $(H_0,H_1)$. The space $\mathcal{A}_{H_0,H_1}$ is of dimension $N_d\times N_d$ and is defined by $\mathcal{A}_{H_0,H_1} = \{M\in \mathbb{R}^{N_d,N_d},
M^\dagger U(t_f) + U^\dagger(t_f)M = 0\}$. The evolution equation of $\delta U$
is obtained by differentiating Eq.~(\ref{eq:dynU}),
\begin{equation}
  \left\{
  \begin{array}{lll}
    i\frac{\partial}{\partial t} \delta U(t) &=& \big[H_0 +H_1E(t)\big]\delta U(t)
  + \big[\delta H_0 +\delta H_1 E(t)\big] U(t)  \\
 \delta U(t=0) &= &0
  \end{array}
    \right. .
  \label{eq:dyndU}
\end{equation}

\subsection{Numerical algorithms}
Before defining the Newton solver, we introduce a relevant time discretization of Eq.~(\ref{eq:dynU}).
\label{subsec:solproblem}
\subsubsection{Crank-Nicholson scheme}
The approach is based on a Crank-Nicholson time discretization of Eq.~(\ref{eq:dynU}).
We give some details about this numerical scheme.  We consider an equidistant time discretization grid
and we denote by
$N_{t_f}$ the number of
sampling points of the time interval $[0,t_f]$,
$dt = t_f/N_{t_f}$ being the time step,
and by $U_n \approx U(t_n)$, $n=0,\cdots,N_{t_f}$, the approximation
of $U$ at a given time grid point $t_n$. The Crank-Nicholson algorithm is based
on the following recursive relation:
\begin{equation}
  i\frac{U_{n+1}-U_n}{dt}  = \big[H_0+E_nH_1\big]\frac{U_{n+1}+U_n}{2} \,,
  \label{eq:CN}
\end{equation}
 with $E_n = E(t_n+\frac{dt}{2})$. Equation (\ref{eq:CN})
can be rewritten in a more compact form as follows:
\begin{equation}
  \big[\openone+L_n\big]U_{n+1}  = \big[\openone-L_n\big]U_n\,,
  \label{eq:CN-Ln}
\end{equation}
where $\openone$ is the identity operator and $L_n = i\frac{dt}{2}(H_0+H_1E_n)$.
This scheme provides a second order approximation
with respect to time, which enables to compute
an accurate approximation of the trajectory $U(t)$.
In addition, note that the Crank-Nicholson propagator preserves the norm.

Differentiating  Eq.~(\ref{eq:CN-Ln}) with respect to ($H_0, H_1$), we
obtain:
\begin{equation}
  \delta L_n \big[U_{n+1}+U_n\big]  = \big[\openone-L_n\big]\delta U_n
  -\big[\openone+L_n\big]\delta U_{n+1}\,,
  \label{eq:CN_deriv}
\end{equation}
where $\delta L_n = i\frac{dt}{2}(\delta H_0+\delta H_1E_n)$.
Combining Eq.~(\ref{eq:CN}) and Eq.~(\ref{eq:CN_deriv}) provides:
\begin{equation}
           U_{n+1}^\dagger\delta U_{n+1}-
           U_{n}^\dagger\delta U_{n} =
          -idt \frac{(U_{n+1}+U_n)^\dagger}{2} \delta L_n \frac{U_{n+1}+U_n}{2}  \,.
  \label{eq:CN_recurs}
\end{equation}
The initial unitary operator $U_0$ being fixed, $\delta U_0
= 0$, and we get from Eq.~(\ref{eq:CN_recurs}),
\begin{equation}
  U_{N_{t_f}}^\dagger\delta U_{N_{t_f}} =
  -idt \sum_{n=0}^{N_{t_f}-1}\frac{(U_{n+1}+U_n)^\dagger}{2} \delta L_n \frac{U_{n+1}+U_n}{2}  \,.
  \label{eq:UNdUN}
\end{equation}
Note that the properties of the Crank-Nicholson scheme is crucial in the latter computation, since
Eq.~(\ref{eq:UNdUN}), which represents the central point of the algorithm
used here, is based on the  Crank-Nicholson relation, i.e. Eq.~(\ref{eq:CN}).

\subsubsection{Newton method}
\label{subsec:newton}
In this section, we define the Newton method used to solve the characterization problem.
Denoting by $\varphi_{dt}(H_0,H_1):=U_{N_{t_f}}$ the time discretized version of $\varphi$ (see Eq.~(\ref{eq:mapphi})) and by
$\bar\varphi_{dt}(H_0,H_1) = \varphi_{dt}(H_0,H_1) - U_\mathrm{tar}$,
the Newton iteration applied to the equation

\begin{equation}\label{eq:EqD}
\varphi_{dt}(H_0,H_1) = U_\mathrm{tar}
\end{equation}
reads:
\begin{equation}
d\bar\varphi_{dt} (H_0^k,H_1^k)\cdot (\delta H_0^k,\delta H_1^k)
= -\bar\varphi_{dt}(H_0^k,H_1^k)
  \label{eq:Newton}
\end{equation}
where $\delta H_0^k$ and $\delta H_1^k$ are the correction terms added to $H_0^k$
and $H_1^k$ at step $k$, to define $H_p^{k+1}
:= H_p^k+\delta H_p^k$ with $p=0,1$.
Since $U_\mathrm{tar}$ is fixed, we deduce that
$$d\bar\varphi_{dt} (H_0,H_1)\cdot (\delta H_0,\delta H_1)
=d\varphi_{dt} (H_0,H_1) \cdot (\delta H_0,\delta H_1).
$$
As in Eq.~(\ref{eq:mapdphi}), the left-hand side of Eq.~(\ref{eq:Newton})
corresponds to $\delta U_{N_{t_f}}^k$ and the latter equation gives rise to:
\begin{equation}
  \delta U_{N_{t_f}}^k = -(\varphi_{dt}(H_0^k,H_1^k) - U_\mathrm{tar}) = U_\mathrm{tar}
  -U_{N_{t_f}}^k.
  \label{eq:UfUtarbar}
\end{equation}
Combining Eq.~(\ref{eq:UNdUN}) and Eq.~(\ref{eq:UfUtarbar}), we obtain
\begin{eqnarray}
  dt \sum_{n=0}^{N_{t_f}-1}\frac{(U_{n+1}^k+U_n^k)^\dagger}{2}
  \big[\delta H_0^k +\delta H_1^k E_n\big] \frac{U_{n+1}^k+U_n^k}{2} \nonumber \\
= i\big[ (U_{N_{t_f}}^k)^\dagger U_{\mathrm{tar}}-\openone\big] \,.
  \label{eq:UNdUNUtar}
\end{eqnarray}

Although the left-hand side of Eq.~(\ref{eq:UNdUNUtar}) is an hermitian operator, its
right-hand side may not be necessary hermitian. The right-hand side is therefore approximated by an hermitian operator, $S^k$,
defined by
\begin{equation}
  S^k := i\frac{(U_{N_{t_f}}^k)^\dagger U_{\mathrm{tar}}
  - U_{\mathrm{tar}}^\dagger U_{N_{t_f}}^k}{2} \,.
  \label{eq:Sk}
\end{equation}
In spite of this approximation, numerical simulations reveal that the Hamiltonians of the system can be determined with a very good accuracy by this approach. The resulting equation is
\begin{eqnarray}
  dt \sum_{n=0}^{N_{t_f}-1}\frac{(U_{n+1}^k+U_n^k)^\dagger}{2}
  (\delta H_0^k +\delta H_1^k E_n) \frac{U_{n+1}^k+U_n^k}{2}= S^k.
  \label{eq:linearedSk}
\end{eqnarray}
Solving Equation (\ref{eq:linearedSk})  with respect to $\delta H_p^k$ with $p=0,1$ requires an inversion of
    linear systems. In view of practical implementations, we rewrite Eq.~(\ref{eq:linearedSk})
    in a more explicit form. By denoting $X_M$ the vector representation
    of a given matrix $M$ in a column-major order, Eq.~(\ref{eq:linearedSk})
    is given by:
  \begin{eqnarray}
    && dt \left(\sum_{n=0}^{N_{t_f}-1}
    \mathcal{H}\bar{U}_n^k \right)X_{\delta H_0^k}
     +dt \left(\sum_{n=0}^{N_{t_f}-1}
    E_n\mathcal{H}\bar{U}_n^k\right) X_{\delta H_1^k}
    = X_{S^k}
    \label{eq:linearedSkvec}
  \end{eqnarray}
  with $$\bar{U}_n^k = \frac{U_{n+1}^k+U_n^k}{2}$$ and



  $$    \mathcal{H}\bar{U}_n    = \bar{U}_n^T\otimes\bar{U}_n^\dagger  , $$
  where $\otimes$ denotes the Kronecker product, and $\bar{U}_n^T$ is the transposed matrix of $\bar{U}_n$.

  The properties of symmetry of $\delta H_0^k$, $\delta H_1^k$ and $S^k$,
induce redundancies in~Eq.~(\ref{eq:linearedSkvec}). The linear system stated in Eq.~(\ref{eq:linearedSkvec}) cannot be directly solved in this form. The  system of scalar equations associated with Eq.~(\ref{eq:linearedSkvec}) is partially
  redundant. Indeed, the equations deal with symmetric matrices (see
  Eq.~(\ref{eq:linearedSk})), so that one shall only consider the scalar equations that
  correspond, e.g., to the entries above the diagonal in
  Eq.~(\ref{eq:linearedSk}).
  Since the matrices $\delta H^k_0$ and $\delta H^k_1$ are
  symmetric, there are also redundancies in the coefficients of the
  unknowns $X_{\delta H_0^k}$ and $X_{\delta H_1^k}$. As a consequence,
  the columns of the matrices involved in Eq.~(\ref{eq:linearedSkvec}) have to be merged (by
  adding) in the case they correspond to the same unknown coefficient of
  $\delta H^k_0$ or $\delta H^k_1$. We denote by
\begin{eqnarray}
     dt \left(\sum_{n=0}^{N_{t_f}-1}
    \mathcal{H}\bar{U}_n^k \right)_{red}X^{red}_{\delta H_0^k}
     +dt \left(\sum_{n=0}^{N_{t_f}-1}
    E_n\mathcal{H}\bar{U}_n^k\right)_{red} X^{red}_{\delta H_1^k}
    = X^{red}_{S^k},
    \label{eq:linearedSkvecRed}
  \end{eqnarray}
the system obtained from Eq.~(\ref{eq:linearedSkvec}) after these reductions. In this reduced form, note that the number of equations is equal to the number of unknowns, i.e. $N_d(N_d+1)/2$.
The Newton algorithm for
the operator characterization problem can then be summarized as follows.
\begin{algorithm}\label{alg:newton}
Given $Tol>0$ and an initial guess $(H_0^0,H_1^0)$
\begin{enumerate}
  \item Set $k=0$ and $e_0=+\infty$.
  \item While $e_k>Tol$ do
\begin{enumerate}
  \item Solve Eq.~(\ref{eq:dynU}) with the Crank-Nicholson propagator~(\ref{eq:CN-Ln}) and the Hamiltonian operators $H_0=H_0^k$ and $H_1=H_1^k$.
  \item Compute the right hand side of Eq.~(\ref{eq:linearedSk}) by using Eq.~(\ref{eq:Sk}).
  \item Reduce the system~(\ref{eq:linearedSkvec}) to get Eq.~(\ref{eq:linearedSkvecRed}).
  \item Compute $X^{red}_{\delta H_p^k}$ with $p=0,1$ the solutions of Eq.~(\ref{eq:linearedSkvecRed}).
\item Define the Hamiltonians for the next iteration $k+1$ by
  $$ H_0^{k+1} = H_0^k + \delta H_0^k $$
  and
  $$ H_1^{k+1} = H_1^k + \delta H_1^k. $$
  \item Set $k=k+1$.
\item Set $e_k=\sum_{p=0,1}\|\delta H_p^{k-1}\|$.
\end{enumerate}
\end{enumerate}
\end{algorithm}

\subsubsection{Continuation method}

\label{subsec:cont}
It is well known that the convergence of the Newton algorithm
is guaranteed only under restrictive conditions
\cite{bryson}. In this way, the method may not converge when the
initial guess is not close enough to the solution. If no accurate approximate solution of
the problem is known, obtaining convergence of the Newton
algorithm can be difficult. To bypass this difficulty, we propose a \emph{globalization strategy} based on a continuation method \cite{bonnard,trelat}.

\begin{figure}[htb]
\centering
    \includegraphics[width=0.9\linewidth]{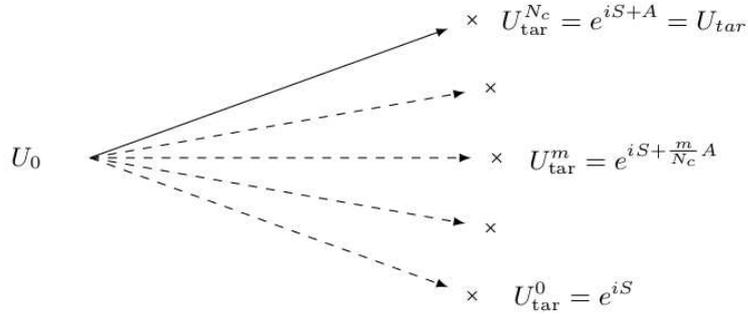}
     \caption{(Color online)
       Decomposition of the target operator into a series of $N_c$ target
       operators with
       $U_{\textrm{tar}}^0 = e^{iS}$,
       $U_{tar}^m = e^{iS+\frac{m}{N_c}A}$, $m=1,\cdots, N_c$
       and $U_{\textrm{tar}}^{N_c} = e^{iS+A} = U_{\textrm{tar}}$.
  }
       \label{fig:U_cib_mov}
\end{figure}

Before summarizing  the continuation method itself, we first
introduce the key idea of the strategy: For a given hermitian operator $U_\mathrm{tar}$, there exist
 symmetric and antisymmetric operators, $S$ and $A$, respectively,
such that
\begin{equation}
  {U}_\mathrm{tar} = e^{iS+A}\,.
  \label{eq:target_sym}
\end{equation}
The matrices $A$ and $S$ can be computed easily by means of a standard eigenvector
solver.
The continuation method consists in solving iteratively operator identification problem by using
intermediate target states defined as:
\begin{equation}
  {U}_\mathrm{tar}^m =
  e^{\frac{m}{N_c}A+iS}\, ,\qquad m=0,\cdots, N_c \,,
  \label{eq:target_inter}
\end{equation}
where $N_c$ is the number of intermediate targets.
Figure \ref{fig:U_cib_mov} illustrates this decomposition of
the target operator into these intermediate targets.
Each step uses the Hamiltonians obtained at the previous step as an initial guess in the Newton solver.

The crucial point is that a solution is known
analytically for the target state with $m=0$. Indeed, the pair $(H_{0,0},H_{1,0}) = (-\frac{S}{t_f},0_{N_d})$ is a solution of
Eq.~(\ref{eq:UtfUtar}) when the right-hand side coincides with $U_{\mathrm{tar}}^0$.
Here $0_{N_d}$ is the $N_d\times N_d$ zero matrix.

We summarize the iterative algorithm
of the continuation method as follows:
\begin{algorithm}
Given $N_c>0$ and set $m=0$, $H_{0,0}=0$, $H_{1,0}=-S/t_f$. While $m<N_c+1$, do
\begin{enumerate}
  \item Define ${U}_\mathrm{tar}^m := e^{\frac{m}{N_c}A+iS}$.
  \item Solve the  operator identification problem by means of Algorithm~\ref{alg:newton} with initial conditions
        $(H_{0,m}^0,H_{1,m}^0)=(H_{0,m-1},H_{1,m-1})$.
  \item Set $m=m+1$.
\end{enumerate}
\end{algorithm}
Note that only the knowledge of the initial and
target operators and the final time, $t_f$, is required.
Through this strategy, the problem is decomposed into a set of smaller problems of the same nature which are solved
sequentially with the Newton algorithm.
\begin{remark}
The pair $(-\frac{S}{t_f},0_{N_d})$ is an exact solution of Eq.~(\ref{eq:UtfUtar}). As a consequence, these matrices may not solve from Eq.~(\ref{eq:EqD}), which includes an approximation, due to time discretization. Instead, one may use
Algorithm~\ref{alg:newton} with $m=0$ to compute accurately a solution associated with the discretized setting and the target ${U}_\mathrm{tar}^0 := e^{iS}$. The pair $(-\frac{S}{t_f},0_{N_d})$ can be used as an initial guess. Moreover, this strategy accelerates  the solving of the step  corresponding to $m=1$ by providing a better initial guess.
\end{remark}
Finally, note that the possible singularities of the Newton method are discussed in the \ref{app}.
\section{Application to quantum systems}
\label{sec:appl}
To test the efficiency of our procedures, we consider the problem of the Hamiltonian identification on two key examples.
\subsection{A driven two-level quantum system}
\label{sec:twolevel}
As a first example, we consider a two-level quantum system.
The Hamiltonian for a two-level atom driven resonantly by a laser
field in the rotating-wave approximation reads \cite{AllenEberly}
\begin{equation}
H(t) =
\left(\begin{array}{cc}
  0 & {\mu}E(t) \\
  {\mu}E(t) & \Delta \\
\end{array}
\right)  \,,
\label{eq:hamtwolevel}
\end{equation}

where $\Delta$ is the detuning term, i.e. the difference between the laser frequency and the frequency of the two-level. The envelope of the control field is assumed to be of the form
\begin{equation}
  E(t) = \frac{1}{2}E_0 \mathcal{E}(t),
\end{equation}
where $\mathcal{E}(t)$ is defined by
\begin{equation}
  \mathcal{E}(t) = \sin^2\left(\frac{\pi t}{t_f}\right)\,.
   \label{eq:shapefunction}
\end{equation}
The strength of the dipole coupling is taken to be $\mu=1\,$a.u. and the
final propagation time is $t_f = 9000\,$a.u.

\subsubsection{Test of the Newton algorithm}
\label{subsec:example1}
The Hamiltonian considered in Eq.~(\ref{eq:hamtwolevel}) can be decomposed as
\begin{equation}
H = H_0 + E(t) H_1
\end{equation}
with
\begin{equation}
{H}_0 =
\left(\begin{array}{cc}
  0 & 0 \\
  0 & \Delta \\
\end{array}
\right)  \,,
{H}_1 =
\left(\begin{array}{cc}
  0 & {\mu} \\
  {\mu} & 0 \\
\end{array}
\right)  .
\label{eq:hamtwolevel2}
\end{equation}
The test of the Newton algorithm is performed as follows:
\begin{enumerate}
\item We consider the Hamiltonian given in Eq.~(\ref{eq:hamtwolevel})
and the initial state
\begin{equation}
{U}_0 =
\left(\begin{array}{cc}
  1 & 0 \\
  0 & 1 \\
\end{array}
\right) \,.
\label{eq:inistate}
\end{equation}
We compute the corresponding final state 
 $U_{N_{t_f}}$ for the given final time $t_f$. The states $U_0$ and $ U_{\mathrm{tar}} = U_{N_{t_f}}$ are respectively chosen as the initial and final states of the Newton algorithm, so that $(H_0,H_1)$ is the solution of the characterization problem.
\item Secondly, we start the Newton algorithm by considering an initial guess of the
form:
\begin{equation}
  \overbrace{H_0+\eta\delta H_0}^{H_0^0} +\overbrace{(H_1+\eta\delta H_1)}^{H_1^0}E(t),
 \label{eq:guessHam}
\end{equation}
where $\delta H_0$ and $\delta H_1$ are chosen randomly with entries in $[-1,1]$.
More precisely, these matrices are determined such that they have
the same symmetry properties as $ H_0$ and $ H_1$, i.e. $\delta H_0$ and $\delta H_1$
belong respectively to $\mathcal{S}^{N_d,N_d}$ and $\mathcal{S}_0^{N_d,N_d}$. The parameter $\eta$ represents the magnitude of the perturbation.
\end{enumerate}
Since for $ U_{\mathrm{tar}} = U_{N_{t_f}}$, the solution ($ H_0,~ H_1$) is known,
the goal is here to analyze the convergence of the Newton procedure with
respect to $\eta$ when the initial guess is given by Eq.~(\ref{eq:guessHam}).
In the following, the detuning is set to $10^{-7}$ a.u.
The convergence of the algorithm is analyzed by averaging over 15 random initial guesses.

We first investigate the convergence of the Newton algorithm
with respect to the number of iterations for a fixed value of $\eta$. At each
iteration of the Newton algorithm, the
deviation from the solution ($ H_0,~ H_1$) is measured by taking the $\log_{10}$
of the norm of the difference between $H_p$ and $H_p^k$, where $k$ is the iteration
index (see Algorithm~\ref{alg:newton}) and $p=0,~1$. The matrix norm is defined here as the maximum absolute value of the eigenvalues of the matrix. In the case $\eta = 10^{-3}$, Table \ref{tab:cong_newton} illustrates the typical quadratic convergence of a Newton algorithm \cite{bryson}.
\begin{table}[htb]
  \centering
   \begin{tabular}{|c| c| c| c| c| c|}\hline
      $k$ &
      $ \log_{10} \| H_0-H_0^k\|$   &
    $ \log_{10} \| H_1-H_1^k\|$   &
     $ \log_{10} \| U_{\mathrm{tar}}-U^k_{N_{t_f}}\|$
      \\ \hline
      1 & -3.5924012132  & -0.9529042109  &  -0.2100714376 \\
      2 & -3.7008867006  & -0.9439569383  &  -1.0991134448 \\
      3 & -3.7090187513  & -0.9508178308  &  -2.6736586007 \\
      4 & -4.8848725497  & -2.1267494911  &  -5.4770040955 \\
      5 & -7.2658303406  & -4.5077075456  &  -7.2259624695 \\
      6 & -11.5917321736 &  -8.8341960474 &  -9.3609642102 \\
      7 & -14.3721229804 & -11.6141682339 & -12.7681301304 \\
      8 & -15.4905077048 & -12.9092171963 & -14.3451244509 \\
      9 & -14.6314758067 & -13.8960673510 & -14.5478814535 \\
      \hline
   \end{tabular}
   \caption{Convergence of the Newton algorithm from $U_0$ to $U_{\mathrm{tar}}$ when the
     guess Hamiltonian is given by Eq.~(\ref{eq:guessHam}), with $\eta = 10^{-3}$.}
  \label{tab:cong_newton}
\end{table}
Starting from a randomly perturbed Hamiltonian, the initial pair
($ H_0,  H_1$) is recovered after few iterations.

\begin{figure}[hbt]
\centering
    \includegraphics[width=0.7\linewidth]{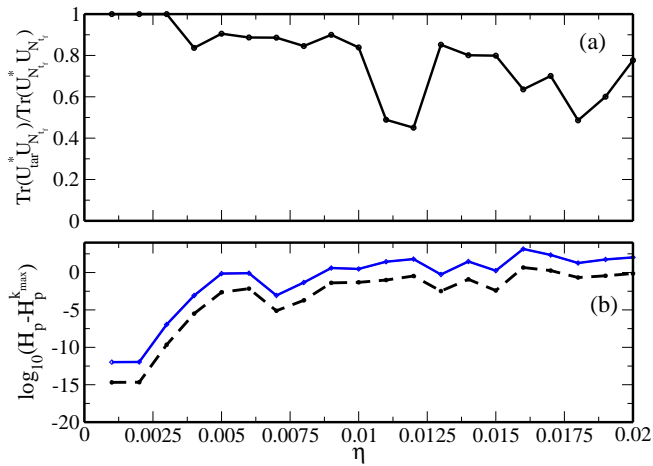}
     \caption{(Color online)
       Convergence of the Newton procedure plotted as a function of the magnitude of the perturbation
       $\eta$. The solid (blue) and the dashed (black) lines of the bottom panel depict respectively the results for the Hamiltonians $H_0$ and $H_1$ at the iteration $k_{\max}= 9$.}
       \label{fig:cong_newton}
\end{figure}

Figure  \ref{fig:cong_newton} shows the convergence behavior of the Newton algorithm as
a function of the parameter $\eta$.
Here again, the accuracy is evaluated by taking the $\log_{10}$ of the norm of the difference between the
Hamiltonians obtained after $k_{\max}=9$
iterations of the algorithm~\ref{alg:newton} and the solution of the problem. As could be expected, the algorithm is efficient for
small values of $\eta$.
In Fig.~\ref{fig:cong_newton}, we observe three cases of convergence behavior:
\begin{enumerate}
  \item Convergence to the initial solution ($H_0,~H_1$) for $\eta \leq 2.10^{-3}$  ($\log_{10} \| H_p-H_p^{k_{\max}}\|<-11$, with $p=0,~1$).
    For small perturbations, the algorithm is able to find the expected solution.
  \item Convergence to another pair of Hamiltonians for $ 2. 10^{-3} \leq \eta \leq 3. 10^{-3} $. It can be seen in Fig.~\ref{fig:cong_newton}
    that the target is reached with
a fidelity close to $100\%$. However, the Hamiltonians found by the algorithm are
not exactly the same as the initial ones since $\log_{10} \| H_0-H_0^{k_{\max}}\| \approx -10$ and
$\log_{10} \| H_1-H_1^{k_{\max}}\| \approx -6$.
This result is a signature of the non uniqueness of the solution.
\item Convergence failure: For larger perturbations ($\eta >3\,10^{-3}$), the Newton algorithm fails to converge.
This may indicate the size of the basin of convergence, which is very narrow.
\end{enumerate}


\subsubsection{Test of the continuation procedure}
\label{subsec:example2}

\begin{figure}[hbt]
\centering
    \includegraphics[width=0.7\linewidth]{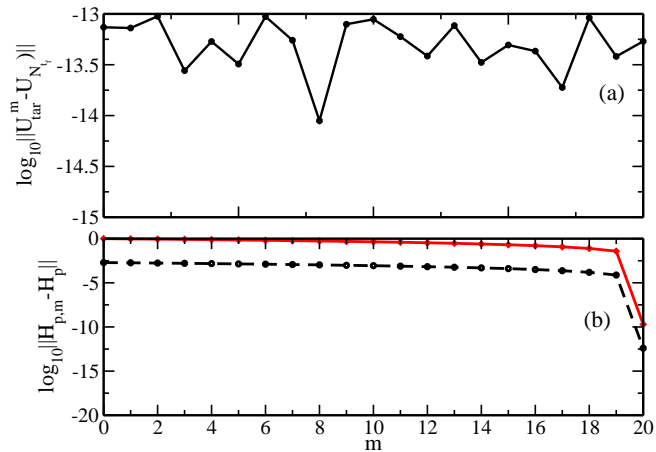}
     \caption{(Color online)
       Convergence of the continuation algorithm plotted as a function of
       the number of iterations, $m$. The target is decomposed into $N_c=20$
       intermediate target operators. The panel (a) corresponds to the deviation of $U_{N_{t_f}}$
      from  $U_\mathrm{tar}^m$ and the panel (b) to the deviation
    $H_{p,m}$ from $H_p$, $p=0,~1$ ($p=0$ black dashed line with circles and $p=1$ red solid line with stars).}
       \label{fig:fig_cont_twolevel}
\end{figure}

In this section, the convergence of the continuation algorithm is tested on the previous two-level quantum system. We keep the same target state $U_\mathrm{tar}$ and the same pair of Hamiltonian solution $(H_0,H_1)$ as in Sec. \ref{sec:twolevel}.
The convergence behavior is illustrated in Fig. \ref{fig:fig_cont_twolevel} with $N_c=20$, the number of intermediate target states.
At each iteration of the continuation algorithm, a very good convergence of the initial state to the
corresponding intermediate target is reached, as can be
seen in Fig.~\ref{fig:fig_cont_twolevel}(a). For each iteration $m$, the
number of required iterations of the Newton algorithm is of the order of 12.
In the lower panel, except for $m=N_c$, the deviation of the
field free and interaction Hamiltonians from the solutions of the problem is  larger than
10$^{-5}$. However, for $m=N_c$, the solution obtained is
approximatively the expected one,
$ \log_{10} \| H_0-H_{0,N_c}\|\approx -12$ and
$ \log_{10} \| H_1-H_{1,N_c}\|\approx -10$.

\subsection{Driven double well potential}
\label{sec:doublewell}
\begin{figure}[htb]
\centering
    \includegraphics[width=0.7\linewidth]{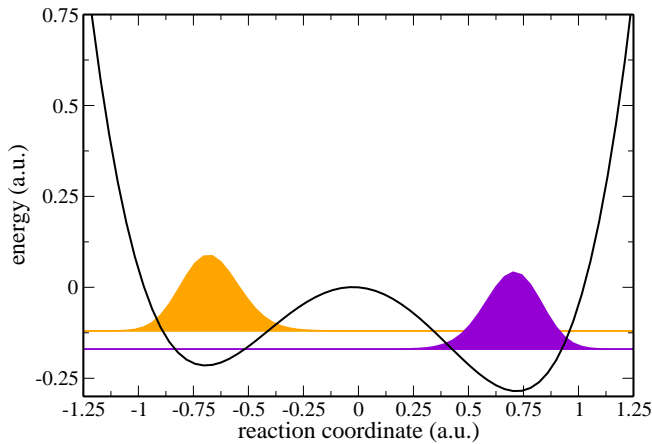}
     \caption{(Color online)
       Potential curve of the molecular system as a function of the reaction coordinate. The probability density of the
     ground state ($v=0$) and of the excited state ($v=3$) are respectively plotted in purple (dark gray) and orange (light gray).}
       \label{fig:pot}
\end{figure}

\begin{figure}[htb]
\centering
    \includegraphics[width=0.7\linewidth]{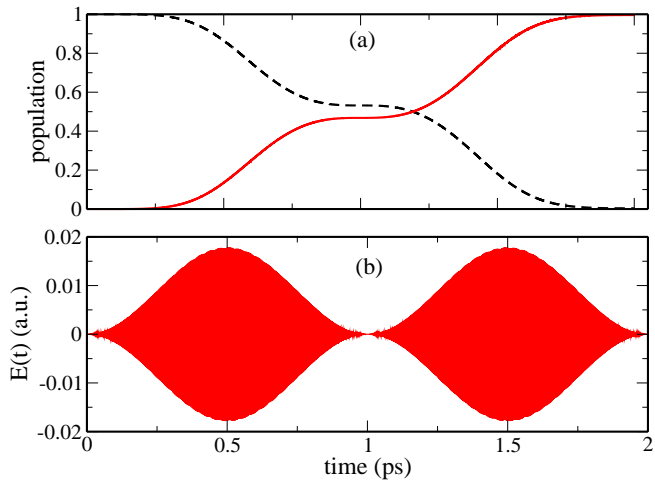}
     \caption{(Color online)
       The top panel represents the time evolution of the population on the states $v=0$ (dashed black
       line) and $v=3$ (red solid line). The bottom part displays the time evolution of the control field.}
       \label{fig:dyn03}
\end{figure}

As a second example, we consider an asymmetric double well potential of
mass $\mathcal{M}= 1000$ a.u.. Figure \ref{fig:pot} displays the energy potential curve of the one-dimensional molecular system used in the computations.
The time-dependent Hamiltonian governing the dynamics of the system is given by
\begin{equation}
  H(r;t) = \underbrace{-\frac{1}{2\mathcal{M}}\frac{\partial^2}{\partial r^2}+r^4 -r^2
  -\frac{1}{20} r}_{H_0} +
  \underbrace{\frac{1}{2} r}_{H_1} E(t) \,,
  \label{eq:hamdw}
\end{equation}
where $r$ is the reaction coordinate and $E(t)$ is the electric field. The final time is set to
$t_f = 2$ ps. To solve the Schr\"odinger equation, we use as basis the eigenvectors of the field-free Hamiltonian $H_0$, which is defined in Eq. (\ref{eq:hamdw}).
We consider the control field that transfers
the population from the ground eigenstate with $v=0$ to the excited eigenstate with $v=3$:
\begin{equation}
  E(t) = \frac{2\pi}{t_f\mu_{0,3}} \sin^2(\frac{4\pi t}{t_f}) \cos(\omega_{0,3}t)\,,
  \label{eq:pipulse}
\end{equation}
where $\mu_{0,3}$ is the matrix element of the coupling
Hamiltonian associated with the eigenstates $v=0$ and $v=3$ and $\omega_{0,3}$ is the energy difference
 between the eigenstates $v=0$ and $v=3$.
Figure~\ref{fig:dyn03} shows the time evolution of the population on the
states $v=0$ and $v=3$ and the corresponding control field.
As in the case of the two-level quantum system, we apply the Newton algorithm by
considering a target operator $ U_{\mathrm{tar}} = U_{N_{t_f}}$ where
$U_{N_{t_f}}$ is the final state obtained with the field of Eq.~(\ref{eq:pipulse}).
We consider a finite Hilbert space of size $N_d= 12$, which is sufficient for
the intensity of the electric field used here.
Setting $\eta = 10^{-6}$, Table \ref{tab:cong_newtondw} illustrates the convergence behavior of the Newton
algorithm.
\begin{table}[htb]
  \centering
   \begin{tabular}{|c| c| c| c| c| c|}\hline
      $k$ &
      $ \log_{10} \| U_{N_{t_f}}-U_\mathrm{tar}^k\|$   &
      $ \log_{10} \| H_0-H_0^k\|$   &
    $ \log_{10} \| H_1-H_1^k\|$
      \\ \hline
    1   &-1.1548028126    &  -5.5961244426   & -2.1831389602   \\
    2   &-3.1579618260    &  -6.3171771473   & -2.1983453573   \\
    3   &-3.1952040375    &  -7.9458560471   & -4.2173783130   \\
    4   &-6.9904587028    & -10.7407123305   & -6.3911179031   \\
    5   &-9.9359223447    & -11.7810549956   & -7.2643429329   \\
    6   &11.4100541376    & -12.2935308612   & -7.6346232913   \\
    7   &11.7710430315    & -12.2099673745   & -7.6105889426   \\
    8   &11.8271010726    & -12.3011508009   & -7.6735489057   \\
    9   &11.8027650635    & -12.2728029350   & -7.6204277008   \\
    10  &-11.9229382795   & -12.2763162205  &  -7.5877013449  \\
    11  &  -11.8190661799 &  -12.2092613111 &  -7.5493066079   \\
      \hline
   \end{tabular}
   \caption{ Convergence of the Newton algorithm from $U_0$ to $U_{\mathrm{tar}}$ when starting from a
     randomly perturbed Hamiltonian, $H_0^0 = H_0+\eta\Delta H_0$ and
     $H_1^0 = H_1+\eta\Delta H_1$ with $\eta = 10^{-6}$.}
  \label{tab:cong_newtondw}
\end{table}
For $\eta = 10^{-6}$, the target is reached with a high accuracy, $ \log_{10} \| U_{N_{t_f}}-U_\mathrm{tar}^{k_{\max}}\| < -11$, with $k_{\max}=11$. However, the Hamiltonian $H_1$ found by the
    algorithm is slightly different from the expected one since
    $\log_{10} \| H_1-H_1^{k_{\max}}\| \approx -7.5$.

 Note that the convergence to the target state is obtained only for very small values of $\eta$ ($\eta<10^{-6}$).
We have also observed that the convergence behavior of the Newton algorithm
depends on the size of the Hilbert space. For example, for $N_d=6$, the Newton algorithm
converges with $\eta=10^{-5}$.

\begin{figure}[hbt]
\centering
    \includegraphics[width=0.7\linewidth]{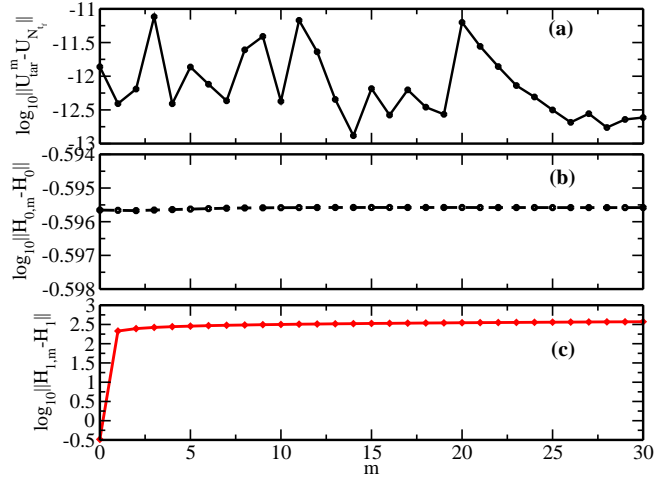}
     \caption{(color online)
       Same as Fig.~\ref{fig:fig_cont_twolevel} but for the example of the
     double well potential.
     The target is decomposed into $N_c=30$
            intermediate target operators
       \label{fig:fig_cont_dw}}
\end{figure}

The convergence and the efficiency of the continuation method are also analyzed
in the case of the double well model for which we have considered the
first $N_d=12$ eigenvectors. For the given initial operator, $U_0 = \openone_{N_d}$, the goal is to identify
a field-free and a coupling Hamiltonians which drive the system to
$ U_{\mathrm{tar}}=U_{N_{t_f}}$ under the
interaction with the electric field.
The target is decomposed into $N_c=30$
intermediate target operators. Each intermediate target state is reached after about 15 iterations of the Newton algorithm.
As in Fig.~\ref{fig:fig_cont_twolevel}, the upper panel of Fig.~\ref{fig:fig_cont_dw} shows
the deviation of $U_{N_{t_f}}$ from  $U_\mathrm{tar}^m$ as a function of
the number of iterations, $m$. Very good results are observed showing the efficiency of the approach. The middle and lower panels of
Fig.~\ref{fig:fig_cont_dw} display
the deviations of the solutions, $H_{0,m}$ and $H_{1,m}$  obtained with the continuation algorithm from
$H_0$ and $H_{1}$, respectively.
At each iteration of the continuation algorithm, an accurate convergence to the intermediate target states is reached. As opposed to the previous example, the Hamiltonian solution obtained
with the continuation method in this case is different from the expected solution.
Here again, this behavior illustrates the non uniqueness of the solution.

\subsection{Discussion}
\begin{figure}[hbt]
\centering
    \includegraphics[width=0.7\linewidth]{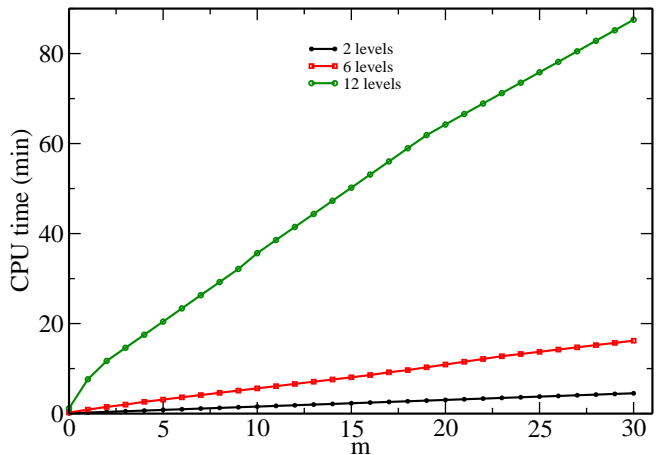}
     \caption{(color online)
       Evolution of the computer time (CPU) in minutes as a function of the number of iterations of the continuation method for the two-level quantum system and for the model molecular one with six and twelve energy levels.
       \label{fig7}}
\end{figure}

We discuss in this final paragraph the numerical cost of the algorithm with respect to the complexity of the quantum system under concern. To illustrate this analysis, we have plotted in Fig. \ref{fig7} the computer time (CPU) used. More precisely, Fig. \ref{fig7} displays the CPU time for the two models as a function of the number of iterations of the continuation method. For a given number of iterations, we see that the computational time roughly increases by a factor 3 from two to six quantum energy levels, that is a quasi-linear increase of the time with the number of levels. However, note that the total duration of the computation is multiplied by a factor 6 when going from 6 to 12 energy levels. This exponential explosion of the CPU time shows that the algorithm cannot be applied actually to more complex systems having, for instance, several dozens of energy levels. The current numerical procedure is therefore too costly and improvements will be required in order to extend this approach to a wider family of quantum systems.
\section{Conclusion}\label{conc}
We have proposed a Newton algorithm to characterize the Hamiltonians of the system in a dynamical setting, i.e. when a control field is applied to the system. A crucial prerequisite of the computational scheme is the perfect knowledge at a given time of the evolution operator of the system. The procedure can be combined with a continuation method in order to enlarge its basin
of convergence. We demonstrate the efficiency of this technique on two key examples, namely a two-level quantum system and a
simple molecular model described by a double-well potential. This work also provides important insights into the different
features of this algorithm such as the size of the basin of the convergence and the non uniqueness of the solution.
Such drawbacks could be removed by considering over-determined data, which is not the case in this paper.
The general applicability of the method makes it an interesting and possibly useful tool to complete techniques of quantum state tomography.

In this work, we have assumed that the evolution operator is perfectly know at a given time without any error. This ideal model has allowed us to highlight the properties of the numerical algorithm.
At this point, in view of applications to more realistic physical systems, the question which naturally arises is the
generalization of this approach to a non-ideal situation in which the propagator can only be estimated to a given accuracy. Another open question would be to consider more complicated quantum systems such as molecular ones with a large number of energy levels. This would require a modification of the present algorithm in order to avoid the explosion of the computational time, as shown in Fig. 6. We are currently working on these open questions.
\section*{Acknowledgments}
Financial supports from the Conseil R\'egional de Bourgogne, the QUAINT coordination action (EC FET-Open) and the
Agence Nationale de la Recherche (ANR), Projet Blanc EMAQS number ANR-2011-BS01-017-01, are gratefully   acknowledged.

\appendix
\section{Singularity problem of the Newton method}\label{app}
 The Newton method is based on the matrix inversion of a Jacobian, and consequently can lead to a numerical explosion if the rank of the latter becomes small. For the continuation method discussed in this paper,
 we can find some examples of intermediate targets for which singularities can occur. We describe such an example. Consider the case of the two-level quantum system presented in Sec.~\ref{sec:twolevel} with a target
 defined by
\begin{equation}
  {U}_\textrm{tar} =
\left(\begin{array}{cc}
  0 & 1 \\
  1 & 0 \\
\end{array}
\right)  \,.
\label{eq:tarapp}
\end{equation}
After decomposition into $N_c$ intermediate targets, the continuation algorithm starts
with the pair $(H_{0,0},H_{1,0}) = (S/t_f,0_{N_d})$, see the Step 1 of the
summary of the continuation method. For the target given by
Eq.~(\ref{eq:tarapp}), the corresponding Hamiltonian $H_{0,0}$ is
\begin{equation}
  H_{0,0} = -\frac{\pi}{2t_f}
\left(\begin{array}{cc}
  1 & -1 \\
  -1 & 1 \\
\end{array}
\right)  =
V^\dagger
\left(\begin{array}{cc}
  -\frac{\pi}{2t_f} & 0 \\
  0 & 0 \\
\end{array}
\right)
V
\,,
\label{eq:H00app}
\end{equation}

where $V$ denotes the matrix of eigenvectors and $V^\dagger$, its adjoint.
At each time, the system is described by
\begin{equation}
  U(t) = e^{-iH_{0,0}t} = V^\dagger
\left(\begin{array}{cc}
  e^{\frac{i\pi t}{2t_f}} & 0 \\
  0 & 1 \\
\end{array}
\right)  V \,.
\label{eq:Utapp}
\end{equation}
In the next step of the continuation algorithm (see Step 2 of the summary of
the continuation method), the Newton algorithm is used to determine $\delta H_{0}$
and $\delta H_1$ from Eq.~(\ref{eq:linearedSk}). In the following, we prove that the Newton algorithm cannot be applied because
of the occurrence of singularities. First, let us recall that the operator identification problem is
equivalent locally to the study of the surjectivity of $\varphi$ of Eq.~(\ref{eq:mapphi}).
As proved in Ref.~\cite{Laurent}, this property  can be
deduced from the surjectivity of $d\varphi$, i.e for a given $S\in
\mathcal{A}_{H_0,H_1}$, the equation
\begin{equation}
  \int_0^{t_f} U^\dagger(t) \big[\delta H0+ \delta H_1E(t)\big] U(t)= S \,,
\label{eq:solSapp}
\end{equation}
has a solution \cite{Laurent}. By inserting Eq.~(\ref{eq:Utapp}) into Eq.~(\ref{eq:solSapp}),
we obtain:
\begin{eqnarray}
  V^\dagger  \int_0^{t_f}
  [
\left(\begin{array}{cc}
  e^{\frac{i\pi t}{2t_f}} & 0 \\
  0 & 1 \\
\end{array}\right)
\delta \tilde{H}_0
\left(\begin{array}{cc}
  e^{-\frac{i\pi t}{2t_f}} & 0 \\
  0 & 1 \\
\end{array}
\right)
\nonumber \\
+
\left(\begin{array}{cc}
  e^{\frac{i\pi t}{2t_f}} & 0 \\
  0 & 1 \\
\end{array}
\right)
\delta \tilde{H}_1 E(t)
\left(\begin{array}{cc}
  e^{-\frac{i\pi t}{2t_f}} & 0 \\
  0 & 1 \\
\end{array}
\right) dt]
  V  = S\,,
\label{eq:DD}
\end{eqnarray}
with $\delta \tilde{H}_p = V^\dagger \delta \tilde{H}_p V$, $p=0,1$. Using the
properties of symmetry of $\delta \tilde{H}_p$, we can express them as:
\begin{equation}
 \delta \tilde{H}_0  =
\left(\begin{array}{cc}
  a & b \\
  b & c \\
\end{array}
\right) \,, \quad
 \delta \tilde{H}_1  =
\left(\begin{array}{cc}
  0 & d \\
  d & 0 \\
\end{array}
\right)
\,.
\label{eq:deltaH0H1}
\end{equation}
Here, we consider that the electric field $E(t)$ is given by Eq.~(\ref{eq:shapefunction}).
With this assumption,
plugging Eq.~(\ref{eq:deltaH0H1}) into Eq.~(\ref{eq:DD}) and evaluating
the integral of this
latter equation, we arrive at:
\begin{equation}
\left(\begin{array}{cc}
  at_f+ct_f & at_f-ct_f-\frac{4ib}{\pi} \\
   at_f-ct_f+\frac{4ib}{\pi} & at_f+ct_f
\end{array}
\right) +
\left(\begin{array}{cc}
  0 & -\frac{14id}{12} \\
  \frac{14id}{12} & 0
\end{array}
\right)
 =  S\,.
\label{eq:DDfinal}
\end{equation}
It is possible to rewrite the left and right hand sides of Eq.~(\ref{eq:DDfinal}) as a vector column:
\begin{equation}
\left(\begin{array}{cc}
  at_f+ct_f \\
  at_f-ct_f-\frac{4ib}{\pi}-\frac{14id}{12} \\
  at_f-ct_f+\frac{4ib}{\pi} +\frac{14id}{12}\\
   at_f+ct_f \\
\end{array}
\right)
 =  X_S\,,
\label{eq:DDvec}
\end{equation}
where $X_S$ denotes the vector representation of the matrix $S$.
We can then rewrite Eq.~(\ref{eq:DDvec}) as:
\begin{equation}
\left(\begin{array}{cccc}
  t_f & 0 & t_f & 0 \\
  t_f & -\frac{4i}{\pi} & -t_f & -\frac{14i}{12} \\
  t_f & \frac{4i}{\pi} & -t_f & \frac{14i}{12} \\
  t_f & 0 & t_f & 0 \\
\end{array}
\right)
\left(\begin{array}{c}
  a\\
  b\\
  c\\
  d
\end{array}
\right)
 =  X_S\,.
\label{eq:DDvecfinal}
\end{equation}
The matrix of the left hand side of Eq.~(\ref{eq:DDvecfinal}) has
a rank equal to 3 and hence is not invertible. Consequently,
the solution of Eq.~(\ref{eq:solSapp}) can not be obtained by
a matrix inversion. We conclude that
the use of the Newton
approach to compute $\delta H_0$ and $\delta H_1$ will lead to singularities in this example.

\end{document}